\documentclass{article}

\usepackage{PRIMEarxiv}

\usepackage[utf8]{inputenc} 
\usepackage[T1]{fontenc}    
\usepackage{hyperref}       
\usepackage{url}            
\usepackage{booktabs}       
\usepackage{amsfonts}       
\usepackage{nicefrac}       
\usepackage{microtype}      
\usepackage{lipsum}
\usepackage{fancyhdr}       
\usepackage{graphicx}       
\graphicspath{{media/}}     
\usepackage{amsmath}
\usepackage{xcolor}
\usepackage{algorithm}
\usepackage{algpseudocode}
\usepackage{booktabs}
\usepackage{tcolorbox}
\title{LMAR: Language Model Augmented Retriever for domain-specific knowledge indexing}

\author{
  Yao Zhao, Yantian Ding, Zhiyue Zhang, Dapeng Yao, Yanxun Xu  \\
  Department of Applied Mathematics and Statistics \\
  Johns Hopkins University \\
  Baltimore\\
}

\begin{document}

\maketitle

\begin{abstract}
Retrieval-Augmented Generation (RAG) systems often struggle with domain-specific knowledge due to performance deterioration of pre-trained embeddings and prohibitive computational costs of large language model (LLM)-based retrievers. While fine-tuning lightweight embedding models offers a promising direction, its effectiveness is limited by the need for high-quality training data and reliable chunking strategies that preserve contextual integrity. We propose LMAR (Language Model Augmented Retriever), a model-agnostic framework that addresses these challenges by combining LLM-guided data synthesis with contrastive embedding adaptation and efficient text clustering. LMAR consists of a two-stage pipeline: (1) Triplet sampling and clustering, where an LLM generates contrastive training data by reasoning over semantic similarities and groups text into coherent units with aligned embeddings; and (2) Question–evidence pair generation, which produces cluster-level supervision signals that align retrieval objectives with broader contextual meaning, where LLMs act as both labeler and validator to ensure high-fidelity supervision throughout the pipeline. Experimental results across multiple domain-specific benchmark datasets demonstrate LMAR outperforms multiple baselines models, while maintaining moderate hardware requirement (7-17 GB VRAM) and low latency (0.13 second). Its model-agnostic nature further enables seamless integration with emerging RAG architectures and text embedding models, ensuring continual improvement without redesigning the pipeline. These results highlight LMAR as a practical and cost-effective solution for scalable domain-specific adaption. LMAR's full implementation can be found at: https://github.com/LMAR2025/LMAR.
\end{abstract}

\section{Introduction}

\begin{figure*}[t]
    \centering
    \includegraphics[width=0.95\textwidth]{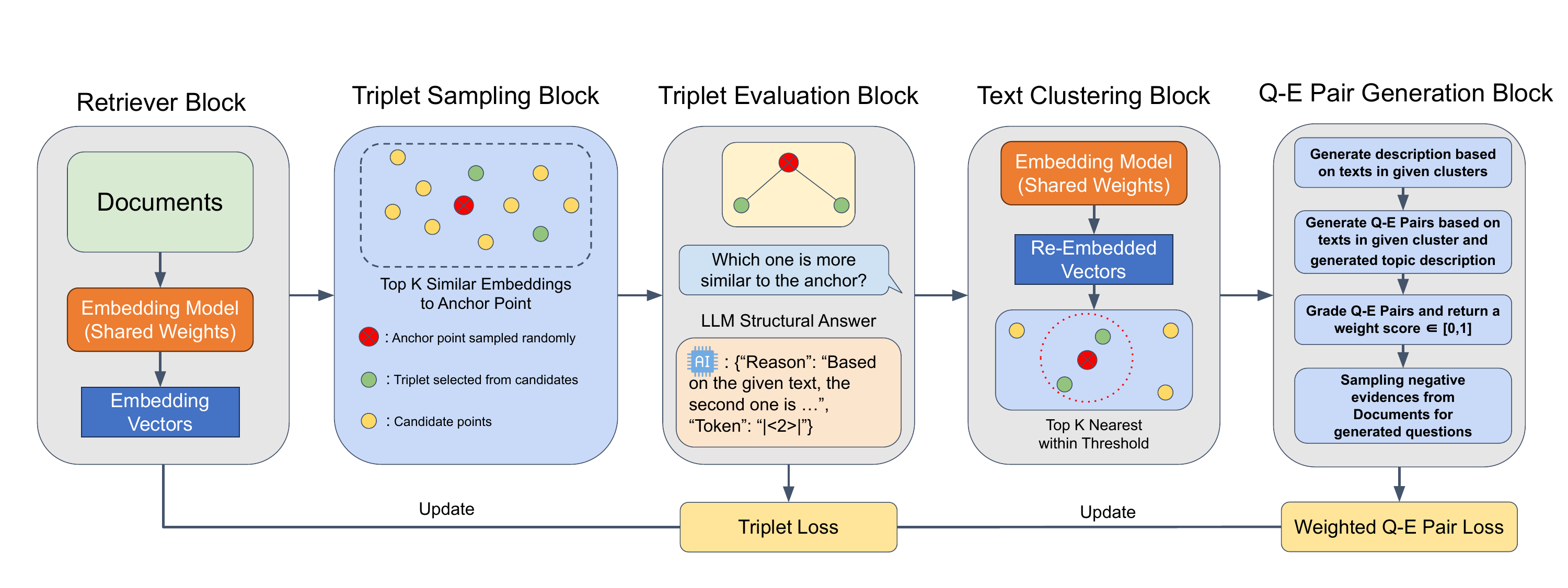}
    \caption{{\bf Overview of the LMAR pipeline}. The five core components: Retriever, Triplet Sampling, Triplet Evaluation, Text Clustering, and Q–E Pair Generation, form a closed-loop training process that refines embeddings and enhances domain-specific retrieval through LLM-guided supervision. }
    \label{fig:RAG pipeline}
\end{figure*}

The integration of Large Language Models (LLMs) into domain-specific applications faces significant challenges across multiple dimensions. At the foundational level, LLMs suffer from inherent limitations including hallucination and factual accuracy issues~\cite{mallen2022not, min2023factscore}, where models generate plausible but incorrect information that can have severe consequences in critical domains such as healthcare, legal, and finance. Additionally, these models exhibit substantial domain knowledge gaps and distribution shift problems~\cite{mecklenburg2024injecting, lin2024flame} when applied to specialized fields, as their training on general corpora often inadequately represents the technical terminology, reasoning patterns, and nuanced knowledge structures required for expert-level performance.
To address the persistent limitations of hallucination and domain knowledge gaps in LLMs, Retrieval-Augmented Generation (RAG)~\cite{lewis2020retrieval} has emerged as a promising solution. By grounding model outputs in external, domain-specific knowledge sources, RAG substantially mitigates hallucination and enables access to specialized information absent from the model’s pre-training data. Although RAG has demonstrated effectiveness~\cite{li2024enhancing, shuster2021retrieval} and is increasingly adopted in practical deployments, its performance and scalability remain constrained by several core challenges. A critical limitation arises from the reliance on indexing text corpus to dense embeddings through embedding models, whose quality deteriorates substantially when applied to out-of-domain corpora. Despite recent advances in developing more powerful LLM-based embedding models~\cite{lee2024nv, solatorio2024gistembed} and increasingly sophisticated RAG architectures~\cite{dong2024advanced, liu2025hm} with promising generalization capability, their adoption in production environments remains constrained by limited computational resources and stringent latency requirements. 

Moreover, text segmentation, a prerequisite step before embedding generation, introduces further complexity. Traditional approaches rely on fixed-length or sentence-based splits that often disrupt contextual coherence, whereas emerging semantic chunking methods aim to partition text based on meaning rather than structure, offering improved contextual integrity. However, the success of semantic chunking depends on the availability of high-quality, domain-adapted embeddings to generate text clusters with high semantic similarity. Conversely, improving embeddings requires supervision that is difficult to construct without reliable, context-aware chunking. This interdependence creates a fundamental bottleneck in the design and deployment of RAG systems for specialized domains and there is no universal solutions proposed to our best knowledge. Recent methods have proposed using LLMs directly to guide chunking decisions \cite{singh2024chunkrag, duarte2024lumberchunkerlongformnarrativedocument, zhao2024meta}, dynamically adjusting chunk boundaries based on discourse structure or learned scoring mechanisms. These approaches, while promising, inherit the limitations of LLMs, particularly hallucination and computational overhead, and can infer misleading topic boundaries or introduce inconsistency under domain shift.

To address these challenges simultaneously and provide a feasible solution for incorporating LLM into domain-specific tasks in industrial scenarios, we propose {\bf Language Model Augmented Retriever (LMAR)}, a practical and scalable framework that jointly optimizes embedding quality and semantic chunking for retrieval-augmented generation while maintaining affordable computation costs and minimum latency. LMAR is based on a two-stage pipeline. In the first stage, LMAR employs LLM-guided data labeling to construct contrastive triplets with hard negative samples, enabling effective embedding refinement through contrastive learning. Then the refined embeddings are used in conjunction with a novel sampling-based K-nearest neighbors (KNN) clustering to form semantically coherent clusters that preserve contextual continuity across text segments. In the second stage, LMAR generates synthetic question–evidence pairs from each cluster, aligning retrieval objectives with the semantic structure of full clusters rather than isolated sentences. This enables context-aware training signals that are better suited to the multi-sentence or paragraph-level reasoning often required in specialized domains. To further improve data fidelity, we propose a novel dual-role validation mechanism in which LLMs function not only as data labelers but also as validators during synthetic data generation. This unified use of LLMs for both supervision and quality control ensures that the generated training data is both semantically accurate and contextually aligned, thereby enhancing the overall reliability and effectiveness of the framework.

Based on these core components, our work offers a model-agnostic pipeline that can be seamlessly integrated with any embedding model or RAG architecture, enabling consistent performance improvements across diverse systems, which is empirically validated through experiments with multiple SOTA embedding models, demonstrating robust gains in domain-specific retrieval tasks. In summary, LMAR offer three key contributions:
\begin{itemize}
    \item \textbf{Model-agnostic Enhancement:} A lightweight framework compatible with universal embedding models and RAG pipelines, enabling consistent improvements without modifying core architectures.
    \item \textbf{Joint Embedding-Chunking Optimization:} A two-stage design that resolves the interdependence between embedding refinement and semantic chunking, producing coherent clusters suited for domain-specific retrieval.
    \item \textbf{LLM-guided Supervision and Validation:} Task-oriented mechanisms leveraging LLMs both for constructing supervision labels and validating synthetic data quality, ensuring high-fidelity training signals without any human labor.
\end{itemize}
These contributions collectively mitigate hallucination and domain knowledge gaps by grounding retrieval in semantically accurate clusters, supporting reliable and resource-conscious deployment in specialized settings.

This paper proceeds as follows. Method section details the LMAR architecture and training procedure. Empirical results presented in Experiments and Results section show that LMAR achieves substantial improvements in retrieval accuracy on domain-specific benchmarks while significantly reducing computational cost compared to LLM-as-embedder baselines. 
Finally, Discussion and Future Work section summarizes its practical advantages and discusses limitations and future directions.

\section{Related Work}
\subsection{Domain Knowledge Indexing}
Early approaches for information retrieval primarily rely on traditional linguistic methods such as BM25~\cite{robertson2009probabilistic} and GloVe~\cite{pennington2014glove} which are based on term frequency, document length normalization, co-occurrence statistics to capture lexical relationships. With recent development of LLMs, recent works have followed two primary directions. One line of research seeks stronger embedding models with enhanced generalization ability. In particular, LLM-based embedding models~\cite{shen2023large, lei2025enhancing, qwen3embedding} have shown substantial gains across public text embedding benchmarks~\cite{enevoldsen2025mmtebmassivemultilingualtext}. Another line of work focuses on developing more sophisticated RAG structure with multi-step retrieval and reasoning ~\cite{trivedi2022interleaving, jiang2023active, jeong2024adaptive, asai2023self, gutiérrez2025hipporagneurobiologicallyinspiredlongterm}. As a tradeoff of improved performances, higher computation cost and latency come along with these new methods. In addition, these methods typically assume access to large-scale annotated training data or extensive domain-specific knowledge, which may not be available in many practical scenarios.

\subsection{LLM-based Data Augmentation}
LLM-based data augmentation are commonly used to address the limitation of constructing high-quality training datasets with sufficient domain coverage and reliable supervision. Recent works have explored LLM-based supervision, where LLMs are used to generate synthetic labels or training data, a process often referred to as LLM distillation~\cite{xu2024survey}. Several strategies have been proposed in this setting, including: (1) using prompt engineering to directly generate data from the model’s internal knowledge through zero-shot or few-shot learning \cite{bonifacio2022inpars, jeronymo2023inpars, boytsov2023inpars, dai2022promptagator, peng2025soft}, (2) generating pseudo queries and annotations at the sentence level \cite{wang2023improving}, and (3) summarizing and generating questions from entire documents \cite{wu2024llm}. Other approaches leverage reinforcement learning techniques \cite{nguyen2024reward}, using LLM as critic model to evaluate and generate reward signals for training retrieval models. While these methods reduce the burden of manual labeling, they often focus on sentence-level supervision or general-topic prompts, overlooking important insights that emerge from synthesizing information across multiple text chunks.

\section{Method}
\label{Methods}
The architecture of LMAR is motivated by the observation that semantic text clustering and retrieval are inherently interdependent: both rely on positioning semantically similar texts close together in the embedding space. However, this circular dependency presents a challenge—high-quality clusters require well-adapted embeddings, and adapting embeddings in turn depends on reliable clusters to guide supervision. To break this loop, LMAR incorporates an LLM as an external supervisor that facilitates both contrastive learning and semantic chunking.

Inspired by recent advances in LLM-based chunking and knowledge distillation, we propose an LLM-augmented text clustering mechanism that distills the contextual understanding and discrimination capabilities of the LLM into a lightweight embedding model. By refining embeddings through contrastive learning and generating clusters that mirror real-world evidence units, LMAR enables the embedding space to evolve toward dual objectives: semantic discrimination for clustering and relevance discrimination for retrieval. This results in a self-reinforcing pipeline where improved embeddings lead to better clusters, which in turn provide higher-quality training signals, ultimately enabling domain adaptation without manual annotation. 

As shown in Figure \ref{fig:RAG pipeline}, the LMAR framework consists of five components that form a closed-loop training pipeline. The {\bf Retriever Block} computes embedding vectors for documents using a shared lightweight encoder. The {\bf Triplet Sampling Block} selects candidate triplets based on similarity to a randomly chosen anchor, which are then passed to the {\bf Triplet Evaluation Block}, where an LLM determines which item is semantically closer to the anchor—providing contrastive supervision via triplet loss. The {\bf Text Clustering Block} uses sampling-based KNN to group the refined embeddings into semantically coherent clusters. Finally, the {\bf Q–E Pair Generation Block} synthesizes questions from each cluster and samples negative evidence to compute a weighted retrieval-aligned question-evidence pair loss. Both contrastive and question-evidence losses are used to iteratively fine-tune the embedding model. Each component contributes to improving the embedding space and retrieval performance. We describe each block in detail below.

\paragraph{Retriver Block} initiates the process by segmenting the input corpus into individual paragraphs, each of which is encoded into an embedding vector using the initial state of the embedding model. Paragraphs are chosen as the fundamental unit of text due to their ability to preserve contextual integrity while balancing granularity and computational efficiency. In contrast, smaller units such as sentences risk often oversimplify complex ideas and fail to capture dependencies across adjacent statements. Sentence-level chunking also produces a large number of segments, making data augmentation and clustering computationally infeasible for large corpora. On the other hand, larger units, such as entire documents or sections, introduce noise by conflating unrelated topics, diluting retrieval precision and frequently exceeding the input length constraints of lightweight embedding models.

\paragraph{Triplet Sampling Block and Triplet Evaluation Block} jointly construct informative anchor-positive-negative triplet training samples to guide contrastive learning and improve the embedding model for downstream clustering tasks. The process begins by selecting a random anchor point from the document embeddings. The top-$K$ most similar candidates are retrieved based on cosine similarity in the embedding space.  From these candidates, two sample points are randomly drawn to form a triplet comprising semantically similar elements in the high dimensional embedding space. These sampled triplets are then evaluated using an LLM to provide supervision for triplet learning by evaluating the semantic similarity within each triplet.   Given an anchor and two candidate samples, the LLM is prompted using a chain-of-thought (CoT) style ~\cite{wei2022chain} to encourage step-by-step reasoning, helping the model differentiate between subtle and challenging sample pairs. The prompts guide the LLM to first analyze and compare key textual elements following a top-down approach starting from topics and tones down to details such as entities and events across the triplet, and then make a judgment on which sample is more semantically similar to the anchor. The output is prompted to generate an explicit reasoning trace that explains the decision before a final answer for labeling. This approach significantly improves the accuracy of labeling task for transformer-based models with autoregressive generation~\cite{sanh2022multitaskpromptedtrainingenables}. When both samples exhibit high similarity to the anchor, the LLM is prompted to return a specific message indicating the ambiguity and skip the labeling process for the current triplet pair. By avoiding cases where genuine semantic similarity exists between all three elements, this strategy prevents false negative labeling that could compromise the contrastive learning objective.

To ensure stable output formatting and facilitate automatic answer parsing, we use prompt-engineered templates with structured outputs and special tokens indicating sample indices. The resulting triplet loss signal is back-propagated to update the embedding model using the triplet margin loss:
\begin{equation*}
L(a, p, n)=\max \left\{d\left(a_i, p_i\right)-d\left(a_i, n_i\right)+\epsilon, 0\right\}
\end{equation*}
where $d\left(x_i, y_i\right)=\left\|\mathbf{x}_i-\mathbf{y}_i\right\|_p$, and $a, p$, and $n$ denote the anchor, positive, and negative samples, respectively, with margin $\epsilon$. 
This LLM-guided supervision enables the embedding model to better align with human-perceived semantic similarity, pulling semantically related texts closer and pushing dissimilar ones farther apart, ultimately improving clustering quality and downstream retrieval performance.

\paragraph{Text Clustering Block} acts as a bridge connecting two different augmentation and fine-tuning tasks. It organizes the corpus into semantically coherent units based on cosine similarity between embedding vectors, reflecting the embedding model's capacity to capture latent relationships within the text for capturing all necessary evidences in retrieval tasks. Unlike traditional text splitting methods that rely on fixed window sizes or syntactic cues, clustering operates directly in the semantic space, allowing it to adaptively group content based on meaning rather than surface structure. This flexibility naturally alleviates issues such as splitting coherent arguments or scattering related evidence across segments, leading to more contextually intact and retrieval-friendly units.

A key challenge in this process is the implementation of traditional clustering methods, such as K-means, to large datasets. In our context, the corpus contains tens of thousands of paragraphs and each cluster is expected to be relatively small to reflect a precise topic, resulting in an enormous number of clusters requiring frequent updates. To address this, our framework proposes an efficient Sampling-Based KNN Clustering, enabling scalable and effective cluster generation even with large datasets and numerous small clusters.

\begin{algorithm}
\caption{Sampling-Based KNN Clustering}
\label{alg:KNN Clustering}
\begin{algorithmic}[1]
    \State \textbf{Input:} Normalized embeddings $\mathcal{Z} \in \mathbb{R}^{n \times d}$, max cluster size $K$, similarity threshold $\delta$
    \State Initialize available indices: $\mathcal{I} \gets \{1, \dots, n\}$
    \State Initialize $\textproc{cluster\_list} \gets \{\}$
    \While{$|\mathcal{I}| > 0$}
        \State Randomly sample $i \sim \mathcal{I}$
        \State $z_i \gets \mathcal{Z}[i]$
        \State $\mathcal{Z}_{\text{avail}} \gets \mathcal{Z}[\mathcal{I}]$
        \State Compute cosine similarities: $s_j = \langle z_i, z_j \rangle$ for $z_j \in \mathcal{Z}_{\text{avail}}$
        \State Select top-$K$ similar embeddings: $\mathcal{S}_K \gets \textproc{TopK}(s_j, K)$
        \State $\mathcal{I}_{\text{cluster}} \gets \{j \mid s_j \in \mathcal{S}_K , s_j > \delta\}$
        \State $\textproc{similarities} \gets \{ s_j \mid j \in \mathcal{I}_{\text{cluster}} \}$
        \State $\textproc{cluster} \gets \{\text{indices}: \mathcal{I}_{\text{cluster}}, \text{similarities}: \textproc{similarities}\}$
        \State Append $\textproc{cluster}$ to $\textproc{cluster\_list}$
        \State Remove $\mathcal{I}_{\text{cluster}}$ from $\mathcal{I}$
    \EndWhile
    \State \textbf{Output: } $\textproc{cluster\_list}$
\end{algorithmic}
\end{algorithm}

As shown in Algorithm~\ref{alg:KNN Clustering}, our method constructs clusters from normalized embeddings by iteratively sampling seed points and grouping their most similar neighbors. At each iteration, the algorithm randomly selects an embedding and computes cosine similarities with all unassigned points. It then forms a cluster by selecting the top K most similar embeddings above a specified similarity threshold \(\delta\). The process continues until all embeddings are clustered. Each iteration computes cosine similarities with remaining points with time complexity of $O(m \cdot d)$, where $m$ is the number of available points and $d$ is the embedding dimension, leading to a worst-case total complexity of $O(n^2 d)$ when all points are compared. The algorithm stores embeddings with space complexity of $O(n d)$ and similarity scores of $O(n)$, and cluster list storage of $O(n)$. This approach emphasizes efficiency and scalability by avoiding full pairwise comparisons through taking out element iteratively, making it suitable for large-scale clustering.

The similarity threshold \(\delta\) is introduced to address the increasing sparsity of the embedding space as clustering progresses, acting as a safeguard against degenerate clusters formed due to random sampling in sparse regions. In the later stages of the algorithm, as more embeddings are assigned to clusters, the remaining unclustered points tend to be more dispersed. As a result, even the top-K most similar neighbors may exhibit low semantic similarity, increasing the risk of forming incoherent clusters. By enforcing a minimum similarity constraint, the algorithm ensures that only embeddings with sufficient semantic alignment are grouped together, hereby maintaining the internal consistency and quality of the resulting clusters.

\paragraph{Question-Evidence Pair Generation Block} employs LLM-based reasoning to synthesize and evaluate question-evidence pairs, thereby aligning the embedding model with task-specific objectives. We utilize the reasoning capabilities of LLM to extract semantically structured information from the text clusters and construct task-specific question-answering data. Each cluster, along with a topic summary generated during earlier reasoning steps, is provided as contextual input to the LLM. The pipeline starts from generating a descriptive summary for topics shared by texts in same clusters, then the model is prompted to generate a set of questions that are tightly aligned with extracted cluster topic, and can only be answered using the associated clustered paragraphs. This design ensures that the generated questions are not only semantically relevant but also grounded in the actual content of the cluster. Subsequently, we evaluate each generated question to determine whether it can be accurately answered using the corresponding evidences and return a score $s \in [0,1]$ acting as the weight to be applied with Q-E Pair Loss for positive samples only, where higher score corresponds to better Q-E dependency. For questions deemed answerable, we pair them with their corresponding clustered paragraph group to form a set of positive question-evidence pairs. These high-quality, semantically coherent pairs collectively constitute a supervised dataset for training. Negative samples are generated by randomly selecting unrelated paragraphs from the entire corpus and pairing them with the original questions to form mismatched question-evidence pairs. The embedding model is then trained to distinguish between semantically aligned and misaligned pairs in the embedding space, yielding a question-evidence pairs loss. Positive and negative pairs are converted to training signal using Cosine Embedding Loss defined as follows: for a pair of embeddings \(x_1, x_2\), a label \(y \in \{1, -1\}\) indicating a positive or negative pair respectively, and a margin \(\epsilon\), the loss is given by \(s*(1 - \cos(x_1, x_2)\) if \(y = 1\), and \(\max(0, \cos(x_1, x_2) - \epsilon)\) if \(y = -1\). This weighted loss is propagated back to the embedding model, providing refinement to the embeddings from the perspective of question-retrieved performance.

To ensure usability for practitioners without prior experience in deep learning or model training, LMAR is designed with a high degree of automation throughout the training pipeline. Components from document segmentation and triplet sampling to LLM-guided evaluation, clustering, and question-evidence pair generation are orchestrated through modular blocks, requiring no manual tuning or labeling. Hyperparameters such as number of neighbors and similarity threshold are dynamically chosen from grid-search for specific-purpose performance, and training progress is automatically monitored using validation loss with early stopping. A small proportion (30\%) of synthetic data is used for validation purpose to ensure the optimal training results. With minimal setup of specifying a directory of raw text documents, users can initiate the full fine-tuning process via a single command-line script. This streamlined design enables industrial users to deploy and adapt LMAR to their corpus without requiring deep technical expertise, lowering the barrier to entry for robust domain-specific RAG systems. Detailed prompts for each components are presented in appendix~\ref{Sec:prompts}.

\subsection{LLM as Data Validator}
We introduce two different mechanisms of employing LLM as data validators for triplet labeling and Q-E pair generation, providing a critical secondary quality control for ensuring the reliability of supervision signals. In the triplet labeling stage, the LLM validator is prompted to assess whether both the positive and negative samples exhibit high similarity to the anchor. If such ambiguity is detected, the validator returns a designated “Error” token, and the triplet is excluded from training. This filtering prevents the inclusion of false-negatives cases where all three samples are semantically related, which could otherwise degrade the effectiveness of contrastive learning. In the Q–E pair generation stage, the validator assigns a confidence score between 0 and 1 to each Q–E pair based on its semantic alignment and contextual coherence. 

We choose these two different mechanisms considering the differing objectives of these two stages. For triplet labeling, a binary filtering mechanism is preferable because contrastive learning depends on a sharp distinction between positive and negative samples by excluding ambiguous triplets altogether preserves clean supervision and ensures stable convergence. By contrast, Q–E pair generation benefits from a weighted approach: synthetic pairs often vary in quality rather than being strictly valid or invalid, and discarding borderline cases would waste potentially informative signals. Using confidence scores as weights allows the model to leverage these nuanced examples proportionally, improving robustness while still prioritizing the highest-quality supervision signals.

\section{Experiments and Results}
\label{Experiments}
Our experiments are designed to validate four core claims about LMAR. First, the necessity of refining embedding models for specialized domains. Second, the effectiveness of LMAR's ability to generalize across diverse and specialized domains without manual data labeling as well as LMAR's model agnostic design that can be applied with any model. Third, the necessity of its clustering components for preserving contextual coherence. Lastly, its practical efficiency under real-world computational and privacy constraints. To assess these claims, we first conduct experiments with several public datasets including WikiQA~\cite{yang2015wikiqa} and TechQA~\cite{castelli2019techqa}. Unlike widely-used benchmarks such as HotpotQA~\cite{yang2018hotpotqa} and TriviaQA~\cite{joshi2017triviaqa}, these two datasets are not in the training corpora of any baseline methods, thereby providing a more rigorous evaluation of true out-of-domain generalization performance. Additionally, we intentionally include PubMedQA~\cite{jin2019pubmedqa} in our evaluation, despite this dataset being present in the training corpora of several baseline models including BGE-M3~\cite{chen2024bge}, Linq-Embed-Mistral~\cite{choi2024linq}, and Qwen3~\cite{qwen3embedding}. This deliberate inclusion aims to validate the performance deterioration of pre-trained embedding models without proper adaptation and to demonstrate LMAR's ability to achieve consistent performance improvements even in domains in which current methods have excels. By testing across these dataset collectively, we prove LMAR’s robustness to domain shifts and its capacity to project high-quality embeddings especially in data-scarce scenarios.

Furthermore, we conduct ablation studies to isolate the impact of clustering to retrieval performance. Specifically, we remove the text clustering components to evaluate whether LLM-generated synthetic data based on standard paragraph chunking can sufficiently support domain adaptation. We also present a representative triplet example from the PubMedQA dataset annotated by the LLM to illustrate how LMAR's triplet labeling and clustering refinement process contributes to improved retrieval accuracy. Finally, we assess LMAR’s flexibility and efficiency through experiments with alternative LLMs. These experiments test how well LMAR maintains performance under resource constraints and demonstrate its adaptability across different model configurations. This dual focus on retrieval accuracy and deployability ensures that LMAR not only advances retrieval performance but also meets the operational requirements of real-world applications with limited GPU availability, data privacy, and low-latency inference. Together, these experiments validate LMAR as a practical solution for domain-specific RAG.

\subsection{Public Dataset Evaluation}
To demonstrate LMAR's effectiveness across different embedding architectures, we conduct public dataset experiments with three base encoder models for LMAR: Sentence-BERT~\cite{reimers2019sentence}, a fine-tuned variant of BERT~\cite{devlin2019bert}, BGE-M3~\cite{chen2024bge}, and Qwen3-Embed-0.6B~\cite{qwen3embedding}, with final embeddings generated through mean pooling of their respective final hidden layer outputs. For LLM-based data augmentation, we use DeepSeek-V3~\cite{liu2024deepseek} as the backbone reasoning model for both triplet evaluation and question-evidence pairs generation. Hyperparameters and training details are reported in Table S1 and Table S2 in Supplementary Section: Training Settings.

We compare the results with three group of baseline models. First group is traditional Linguistic Models including BM25~\cite{robertson2009probabilistic} and GloVe~\cite{pennington2014glove}. These models focus on lexical matching and statistical co-occurrence patterns rather than semantic understanding. Second group includes parameter-efficient language models: Sentence-BERT (0.06 B Parameter Size), BGE-M3 (0.5 B Parameters), and Qwen3-Embed-0.6B (0.6B Parameter Size). This category represents the practical deployment scenario where computational resources are limited but semantic understanding is still required. These models are particularly relevant for evaluating LMAR's effectiveness in resource-constrained environments typical of real-world RAG deployments. Lastly, we compare the results against SOTA open-source LLM-based embedding model highest-ranked on the MTEB Leaderboard~\cite{enevoldsen2025mmtebmassivemultilingualtext} including Qwen3-Embed-8B (8B Parameter Size) and Linq-Embed-Mistral (8B Parameter Size) serving as upper-bound baselines to assess whether LMAR's domain adaptation approach can achieve competitive performance against significantly larger models.

We evaluate retrieval performance using four metrics: accuracy, mean reciprocal rank (MRR), average similarity, and term frequency overlap score (TF score), with top five retrieved evidences. These metrics collectively assess task performance, ranking quality, embedding-level semantic alignment, and lexical coverage. Accuracy is calculated as the proportion of question-evidence pairs in which the true evidence appears in the top five retrieved results. MRR measures the average inverse rank of the first relevant retrieval across all evaluation queries, providing insight into how well the retriever prioritizes correct results. To evaluate the semantic effectiveness of the fine-tuned embedding model, we compute the average similarity score between each question-evidence pair using cosine distance in embedding space, both before and after fine-tuning. An increase in similarity indicates that fine-tuning has aligned semantically related question-evidence pairs more closely, improving the model’s ability to retrieve relevant and accurate text segments. Since BM25 model doesn't apply embedding vectors for retrieval, average similarity score is not applicable for it.

TF score is introduced to assess lexical alignment between retrieved paragraphs and their corresponding ground-truth evidence. 
Since retrieved paragraphs are generally longer than the evidence texts, this metric evaluates how well the key lexical content of the evidence is emphasized in the retrieval. Specifically, both the evidence \(e\) and the retrieved content \(r\) are first transformed into TF vectors, where each entry represents the absolute frequency of a term within the respective text. Given a vocabulary of size \(N\), the TF score is computed as:
\begin{equation*}
\operatorname{TFScore}(e, r) = \frac{\sum_{i=1}^{N} \mathrm{TF}_e(i) \cdot \mathrm{TF}_r(i)}{\sum_{i=1}^{N} \mathrm{TF}_r(i)}.
\end{equation*}
This score yields a weighted summation of evidence term frequencies, where each term is scaled by its corresponding frequency in the retrieved documents. A higher score indicates better lexical alignment. The denominator normalizes the contribution by the overall term usage in the retrieved content, penalizing overly verbose paragraphs and favoring concise, targeted matches.

\begin{table}[h]
\centering
\caption{Performance comparison of LMAR and baseline models on public datasets. The best result in each metric is bolded.}
\begin{tabular}{cccccc}
\hline \hline
Model Type                       & Accuracy          & MRR             & TF Score          & Average Similarity \\ \hline \hline

\multicolumn{5}{l}{\textbf{Dataset: WikiQA}} \\ \hline
BM25                                   & 0.61              & 0.48            & 2.05             & -               \\
GloVe                                  & 0.63              & 0.45            & 2.15             & 0.70               \\
S-BERT                                  & 0.69              & 0.51            & 2.06              & 0.47               \\
BGE-M3                      & 0.82 & 0.67 & 2.33  & 0.67 \\
Qwen3-Embed-0.6B             & 0.71              & 0.50            & 2.11              & \textbf{0.83} \\
Linq-Embed-Mistral                    & 0.56              & 0.38            & 1.89              & 0.56               \\
Qwen3-Embed-8B             & 0.49              & 0.36            & 1.82              & 0.82               \\
LMAR (S-BERT)     & 0.74     & 0.55   & 2.18     & 0.59    \\
LMAR (BGE-M3)     & \textbf{0.87} & \textbf{0.69} & \textbf{2.41} & 0.50  \\
LMAR (Qwen3-Embed-0.6B)      & 0.85     & 0.63   & 2.37     & 0.59      \\ \hline
\multicolumn{5}{l}{\textbf{Dataset: TechQA}} \\ \hline
BM25                                   & 0.50              & 0.44            & 28.21             & -               \\
GloVe                                  & 0.47              & 0.39            & \textbf{30.86}    & \textbf{0.89}               \\
S-BERT                                  & 0.52              & 0.42            & 14.02             & 0.61               \\
BGE-M3                      & 0.67 & 0.58 & 7.35 & 0.64 \\
Qwen3-Embed-0.6B             & 0.83              & 0.75            & 19.83              & 0.87 \\
Linq-Embed-Mistral                     & 0.70              & 0.61   & 13.30             & 0.69      \\
Qwen3-Embed-8B             & 0.84              & 0.75            & 21.36              & 0.83               \\
LMAR (S-BERT)                  & 0.71     & 0.61   & 15.76    & 0.52               \\
LMAR (BGE-M3)                & 0.77 & 0.67 & 17.74 & 0.40 \\
LMAR (Qwen3-Embed-0.6B)      & \textbf{0.85}     & \textbf{0.76}   & 18.55     & 0.70      \\ \hline

\multicolumn{5}{l}{\textbf{Dataset: PubMedQA}} \\ \hline
BM25                                   & 0.82              & 0.74            & 3.45             & -               \\
GloVe                                  & 0.75              & 0.67            & 3.02             & 0.78               \\
S-BERT              & 0.84              & 0.73            & 2.40              & 0.53               \\
BGE-M3            & 0.95  & 0.94  & 3.52  & 0.61 \\
Qwen3-Embed-0.6B             & 0.98              & 0.97            & 3.70              & \textbf{0.83}               \\
Linq-Embed-Mistral                     & 0.96     & 0.92   & 2.84              & 0.63               \\
Qwen3-Embed-8B             & 0.98              & \textbf{0.97}            & \textbf{3.94}              & 0.78               \\
LMAR (S-BERT)                                  & 0.95              & 0.89            & 3.25     & 0.55               \\
LMAR (BGE-M3)    & 0.98              & 0.96           & 3.67      & 0.30               \\
LMAR (Qwen3-Embed-0.6B)      & \textbf{0.99}     & 0.96   & 3.74     & 0.74      \\ \hline
\hline \hline
\end{tabular}
\label{table:Main-Result-table}
\end{table}

The results on the public datasets are summarized in Table \ref{table:Main-Result-table}. These results provide strong support for the effectiveness and necessity of fine-tuning lightweight embedding models for domain-specific tasks. On the PubMedQA dataset, Qwen3-Embed-8B and Linq-Embed-Mistral show almost perfect accuracy showing their strong capability on pretrained datasets and domains. While the LLM-based baseline models ranks highly on benchmark leaderboard with outstanding performance on pretrained datasets, they both underperform on WikiQA, falling behind the parameter-efficient baseline models and even the linguistic models. In addition, parameter-efficient baseline models both shows a more robust performance across different datasets indicating stronger generalization capabilities. This highlights the critical limitation of LLM-based embedding models with enormous parameter sizes as they often lack robustness when transferred to domains or formats that diverge from their training distributions. Despite strong average performance across multiple tasks, such models can struggle with fine-grained distinctions or lexical precision in specific applications.

Based on the comprehensive evaluation results across three distinct datasets, LMAR consistently achieves superior performance with accuracy scores of 0.87, 0.85, and 0.99 respectively. Its domain-adaptive capability is validated particularly in the WikiQA dataset, where LMAR (BGE-M3) achieves the highest accuracy given most baselines shows major performance deterioration, its effectiveness even in highly specialized fields that typically require extensive domain expertise for manual annotation. Notably, LMAR's model-agnostic design is evidenced by the consistent improvements compared to corresponding baseline models serving as LMAR's base encoder (S-BERT, BGE-M3, Qwen3-Embed) across three datasets, indicating that the approach can be effectively applied regardless of the underlying embedding model choice.

\subsubsection{Training Effect on Embedding Space}
One important result deviated from our expectation is that average Cosine similarity between evidence-question pairs decreases instead of increases after finetuning the embedding model on some datasets, despite improvements in Accuracy, MRR, and TF Score. We attribute this phenomenon to these key factors:

First, LMAR employs contrastive learning with triplet margin loss to enforce a relative separation between positive and negative samples rather than maximizing absolute similarity. By prioritizing a margin between positive and negative pairs, the model focuses on improving ranking discrimination. This often results in a compressed similarity distribution where both positive and negative pairs occupy lower absolute similarity scores, yet maintain a critical separation threshold. While this reduces average similarity across all pairs, it enhances the model’s ability to rank positives above negatives—directly benefiting ranking-dependent metrics like MRR and Accuracy.

Furthermore, the cosine embedding loss also explicitly penalizes negative pairs by pushing their similarity below a margin, often leading to a downward shift in the overall similarity distribution. When combined with large proportion of negative samples, which is 4 times the size of positive pairs in LMAR, the model allocates significant capacity to separate negatives, further reducing average similarity. The baseline model, lacking such fine-tuning, retains generic semantic embeddings with higher average similarity but poorer task-specific discrimination.

Finally, metrics such as Accuracy and MRR evaluate relative performance, while average similarity measures absolute embedding proximity. The former are agnostic to absolute similarity values as long as the ranking order is preserved. For instance, even if all pairwise similarities decrease, a preserved ranking hierarchy will still yield higher MRR. Conversely, TF Score—which measures lexical overlap—aligns with the model’s domain-specific fine-tuning to prioritize evidence coverage, independent of embedding similarity trends.

\subsection{Ablation Study}
\subsubsection{Triplet Labeling and Clustering Structure}
To assess the contribution of the clustering components in our LMAR framework, we conduct an ablation study by removing the text clustering components, while keeping all other components and training settings unchanged. This variant, denoted as LMAR No Cluster, represents a standard fine-tuning pipeline on LLM-synthetic data without clustering guidance. As shown in Table~\ref{table:Ablation-Result-table}, the removal of clustering components consistently degrades retrieval performance. These results demonstrate that LLM-guided clustering is a necessary step for fine-tuning embedding models in new domains and significantly enhances retrieval quality by enforcing semantic coherence during both the embedding training and text chunking stages. In particular, datasets like TechQA and PubMedQA benefit more than WikiQA from clustering-based augmentation, likely due to their longer and more specialized context passages. For instance, TechQA often includes multi-step technical solutions that require preserving the continuity of semantically linked instructions. Without clustering, synthetic question-evidence pairs risk fragmenting key evidence, leading to lower retrieval accuracy.

\begin{table}[ht!]
\centering
\caption{Ablation study comparing LMAR (S-BERT) with and without clustering components.}
\begin{tabular}{cccccc}
\hline \hline
Model Type         & Accuracy          & MRR             & TF Score          & Average Similarity \\ \hline \hline
\multicolumn{5}{l}{\textbf{Dataset: WikiQA}} \\ \hline
LMAR               & \textbf{0.74}              & \textbf{0.55}            & \textbf{2.18}              & \textbf{0.59}               \\
LMAR No Cluster    & 0.71              & 0.48            & 1.97              & 0.35               \\ 
\hline

\multicolumn{5}{l}{\textbf{Dataset: TechQA}} \\ \hline
LMAR               & \textbf{0.71}              & \textbf{0.61}            & \textbf{15.76}             & \textbf{0.52}               \\
LMAR No Cluster    & 0.68              & 0.59            & 13.44             & 0.46               \\ 
\hline

\multicolumn{5}{l}{\textbf{Dataset: PubMedQA}} \\ \hline
LMAR               & \textbf{0.95}             & \textbf{0.86}              & \textbf{3.25}              & 0.55               \\
LMAR No Cluster    & 0.87             & 0.79              & 3.15              & \textbf{0.59}               \\ \hline \hline
\end{tabular}
\label{table:Ablation-Result-table}
\end{table}

To further illustrate why clustering refinement is essential, we examine the underlying mechanism through which triplet sampling addresses semantic similarity challenges and present a representative example triplet case from PubMed dataset. In this pediatric fracture diagnosis example~\ref{examplecase}, the anchor text discusses comparing ultrasound and X-ray methods, while both candidate examples address the same general medical topic. However, the negative example initially exhibits a higher similarity score (0.84) than the positive example (0.78) despite being semantically less aligned with the anchor's comparative research objective. This occurs because the negative example contains more overlapping keywords and statistical details such as "162 of 248 bones," "130 fractures," and "148 using X-ray" that create superficial lexical similarity, whereas the positive example provides the conceptual conclusion that directly answers the research question posed in the anchor. Without triplet loss refinement, the clustering algorithm would incorrectly group the anchor with the negative example due to higher initial similarity and exclude the positive evidence that contains the essential conclusion. Through triplet loss training, the similarity scores are appropriately adjusted—the positive example's similarity increases to 0.91 while the negative example's decreases to 0.66 ensuring semantically coherent and contextually relevant texts are positioned closer in the embedding space, ultimately forming clusters that preserve the logical flow from research questions to their corresponding answers.

\subsubsection{LLM Reasoning Models}
We also evaluate LMAR’s adaptability across different backbone LLM reasoning models. First, we compare the performance of three LLMs used in the reasoning components of LMAR: DeepSeek-V3 \cite{liu2024deepseek}, GPT-4o \cite{openai2024gpt4o}, and LLaMA3.1-8B-Instruct \cite{touvron2024llama3}.
Detail results are presented in Table~\ref{table:Base-LLM-table}. While GPT-4o achieves the highest overall metrics, Llama3.1-8B still delivers competitive results across all datasets, highlighting LMAR’s effectiveness even when paired with lightweight LLMs. This demonstrates LMAR’s flexibility and suitability for deployment under various computational and privacy constraints.

\begin{center}
    \begin{tcolorbox}[
        colback=blue!5,
        colframe=blue!40!black,
        boxrule=2pt,
        arc=4pt,
        left=12pt,
        right=12pt,
        top=12pt,
        bottom=12pt,
        title={\Large\textbf{Triplet Sampling Example Case}},
        coltitle=white,
        colbacktitle=blue!80!black,
        fonttitle=\bfseries
    ]
    
    \begin{tcolorbox}[
        colback=gray!10,
        colframe=gray!60,
        boxrule=1pt,
        arc=2pt,
        left=8pt,
        right=8pt,
        top=8pt,
        bottom=8pt,
        title={\textbf{Corresponding Question}},
        coltitle=black,
        colbacktitle=gray!30
    ]
    Is ultrasound equal to X-ray in pediatric fracture diagnosis?
    \end{tcolorbox}

    \vspace{8pt}
    
    \begin{tcolorbox}[
        colback=yellow!10,
        colframe=yellow!60,
        boxrule=1pt,
        arc=2pt,
        left=8pt,
        right=8pt,
        top=8pt,
        bottom=8pt,
        title={\textbf{Anchor Text}},
        coltitle=black,
        colbacktitle=yellow!30
    ]
    Ultrasound is currently not established for the diagnosis of fractures. The aim of this study was to compare ultrasound and X-ray beyond their use solely for the identification of fractures, i.e., for the detection of fracture type and dislocation for pediatric fracture diagnosis.
    \end{tcolorbox}
    
    \vspace{8pt}
    
    \begin{tcolorbox}[
        colback=green!10,
        colframe=green!60,
        boxrule=1pt,
        arc=2pt,
        left=8pt,
        right=8pt,
        top=8pt,
        bottom=8pt,
        title={\textbf{Positive Evidence}},
        coltitle=black,
        colbacktitle=green!30
    ]
    \textbf{Text:} Thus, ultrasound can be used as an adequate alternative method to X-ray for pediatric fracture diagnosis.
    
    \vspace{6pt}
    \textbf{Similarity Score with Anchor:} \colorbox{green!20}{\textcolor{green!70!black}{\textbf{0.78 → 0.91}}} \textcolor{green!70!black}{\textbf{(+0.13 improvement through Triplet Loss Finetuning)}}
    \end{tcolorbox}
    
    \vspace{8pt}
    
    \begin{tcolorbox}[
        colback=red!10,
        colframe=red!60,
        boxrule=1pt,
        arc=2pt,
        left=8pt,
        right=8pt,
        top=8pt,
        bottom=8pt,
        title={\textbf{Negative Evidence}},
        coltitle=black,
        colbacktitle=red!30
    ]
    \textbf{Text:} 162 of 248 bones were fractured. 130 fractures were identified using ultrasound, and 148 using X-ray. There were some advantages of X-ray over ultrasound in the detection of fracture type (80 correct results using X-ray, 66 correct results using ultrasound). Ultrasound, however, was superior to X-ray for dislocation identification (41 correct results using X-ray, 51 correct results using ultrasound).
    
    \vspace{6pt}
    \textbf{Similarity Score with Anchor:} \colorbox{red!20}{\textcolor{red!70!black}{\textbf{0.84 → 0.66}}} \textcolor{red!70!black}{\textbf{(-0.18 decrease through Triplet Loss Finetuning)}}
    \end{tcolorbox}
    \end{tcolorbox}
    \label{examplecase}
\end{center}


\begin{table*}[h]
\centering
\caption{Evaluation of different LLM models (DeepSeekV3, GPT-4o, and Llama3.1-8B) with LMAR (S-BERT).}
\begin{tabular}{lccccc}
\hline \hline
Model Type                        & Accuracy & MRR             & TF Score          & Average Similarity \\ \hline \hline
\multicolumn{5}{l}{\textbf{Dataset: WikiQA}} \\ \hline
LMAR-DeepSeekV3                        & \textbf{0.74}     & \textbf{0.55}            & 2.18              & 0.59      \\
LMAR-GPT4o                             & \textbf{0.74}     & 0.54            & \textbf{2.32}             & \textbf{0.65}      \\
LMAR-Llama3.1-8B                       & 0.70     & 0.51            & 2.19             & 0.59      \\
\hline
\multicolumn{5}{l}{\textbf{Dataset: TechQA}} \\ \hline
LMAR-DeepSeekV3                        & 0.71     & 0.61            & \textbf{15.76}             & 0.52      \\
LMAR-GPT4o                             & \textbf{0.73}     &\textbf{0.62} & 12.89             & \textbf{0.57}      \\
LMAR-Llama3.1-8B                       & 0.66     & 0.57            & 15.56             & 0.53      \\
\hline
\multicolumn{5}{l}{\textbf{Dataset: PubMedQA}} \\ \hline
LMAR-DeepSeekV3                & \textbf{0.95}     & 0.86            & 3.25              & 0.55               \\
LMAR-GPT4o                     & 0.92              & \textbf{0.87}   & \textbf{3.43}     & \textbf{0.61}      \\
LMAR-Llama3.1-8B               & 0.88              & 0.78            & 2.54              & 0.39               \\
\hline \hline
\end{tabular}
\label{table:Base-LLM-table}
\end{table*}

\subsection{Model Efficiency Evaluation}
\label{Sec:Model Efficiency Eval}
To further demonstrate LMAR’s efficiency and scalability, we evaluate its API token consumption, hardware usage, and latency. For API consumption evalutiona, we introduce a metric, Token Consumed per Document Token (TCDT), to quantify the efficiency of LLM-based synthetic data generation. TCDT is defined as the total number of tokens consumed, i.e., the sum of input and output tokens normalized by the number of document tokens processed. This metric provides a direct measure of token expenditure across datasets of varying sizes, and is valuable for evaluating the scalability of retrieval-augmented frameworks like LMAR. From Table~\ref{table:Token-Usage-Table} we observe that on the TechQA dataset, LMAR consumes 6.13 Million tokens (TCDT ratio: 6.25), and the ablated model uses only 1.19M tokens (TCDT: 1.21), reflecting the additional overhead from clustering and triplet sampling. Similar patterns are observed from WikiQA and PubMedQA datasets. Even with higher LLM reasoning cost, the overall efficiency of LMAR still support it for all kinds of real world retrieval tasks. In our experiments, LMAR exhibits a TCDT ratio of approximately 6X, meaning that for each document token processed, about six tokens are consumed by the LLM API. This offers a practical estimate of the implementation budget required for synthetic data generation, based solely on document token counts.

\begin{table}[H]
\centering
\caption{Token usage comparison between LMAR models with and without clustering on TechQA and PubMedQA. Clustering significantly increases token consumption due to expanded input queries, but yields substantial gains in retrieval effectiveness. TCDT (Token Consumed per Document per Token) quantifies the token efficiency of each configuration.}
\resizebox{\textwidth}{!}{
\begin{tabular}{llccccc}
\hline \hline
Dataset & Model Type & Document Token & Input Token & Output Token & Total Token Used & TCDT \\
\hline
TechQA & LMAR & 979{,}843 & 5{,}883{,}005 & 243{,}489 & 6{,}126{,}494 & 6.25 \\
TechQA & LMAR No Cluster & 979{,}843 & 1{,}076{,}347 & 112{,}166 & 1{,}188{,}513 & 1.21 \\
PubmedQA & LMAR & 355{,}886 & 1{,}701{,}065 & 390{,}632 & 2{,}091{,}697 & 5.88 \\
PubmedQA & LMAR No Cluster & 355{,}886 & 428{,}994 & 145{,}452 & 574{,}446 &1.61 \\
WikiQA& LMAR & 855{,}397& 5{,}092{,}814& 297{,}009& 5{,}389{,}823 &6.30\\
WikiQA& LMAR No Cluster & 855{,}397 & 998{,}630& 327{,}235 & 1{,}325{,}865 & 1.55\\
\hline
\end{tabular}
}
\label{table:Token-Usage-Table}
\end{table}

Next we present the hardware computational cost for implementing LMAR system in Table~\ref{table:Training-Time-Table}. All training are conducted on one single A100 card with 40G VRAM. For TechQA, LMAR with SENTENCE-BERT uses 8GB VRAM and requires 5.5 minutes (0.3 minutes triplet training + 5 minutes QE-pair training). BGE-M3, a larger model (17GB VRAM required), achieves significantly slower triplet training (3 minutes) and QE-pair training time (40 minutes) due to its parameter scale. PubMedQA training follows a similar pattern. The balance between efficiency and performance highlights LMAR's strength—delivering domain-specific accuracy with modest hardware demands while retaining the capability to integrate more powerful models when additional computational resources are available. This makes it an optimal choice for both lightweight local deployment and high-performance retrieval systems. 

For fully local deployment, the data augmentation and training process of LMAR requires 7 to 17 GB of GPU VRAM, depending on the choice of base embedding models and LLMs used for reasoning. Notably, integrating LMAR with Llama3.1-8B model for chat-based retrieval can be accomplished with approximately 7.5 GB of VRAM when using appropriate quantization techniques, making it well-suited for local implementation on widely available consumer-grade GPUs.

\begin{table}[H]
\centering
\caption{GPU memory usage and training time (in minutes) for different LMAR variants. The full LMAR pipeline requires GPU resources (7–17 GB) and additional time for triplet and QE pair training. Removing the clustering structure reduces training time significantly but sacrifices retrieval quality, highlighting a trade-off between computational efficiency and model performance.}

\begin{tabular}{llccc}
\hline \hline
Base Model & Model Type & GPU VRAM & Triplet Train & QE Pair Train \\ \hline
SENTENCE-BERT & LMAR & 7G & 3 & 5 \\
BGE-M3 & LMAR & 17G & 0.3 & 40 \\
SENTENCE-BERT & LMAR no Cluster & 7G & - & 3 \\
BGE-M3 & LMAR no Cluster & 17G & - & 40 \\
Ling-Embed-Mistral & Deploy Only & 13G & - & - \\
\hline \hline
\end{tabular}
\label{table:Training-Time-Table}
\end{table}

\begin{table*}[h]
\centering
\caption{Latency test results for baseline models and three variants of LMAR. The response time is corresponding to the embedding and retrieval time with one single test input.}

\begin{tabular}{cc}
\hline \hline
Model         & Response Time (second) \\ \hline
BM25                 &  0.004 \\
GloVe               &  0.002 \\
Linq-Embed-Mistral   &  0.54 \\
Qwen3-Embed-8B       &  0.52 \\
\textbf{LMAR (S-BERT)} & \textbf{0.13} \\
LMAR (BGE-M3) & 0.27\\
LMAR (Qwen3) & 0.31\\
\hline \hline
\end{tabular}
\label{table:Training-Latency}
\end{table*}

LMAR's responding time for one single retrieval entry is reported in Table~\ref{table:Training-Latency}. LMAR (S-BERT) obtains lowest latency time of 0.13 second compared to 0.54 second for LLM-based embedding model (Linq-Embed-Mistral). One of the most importance advantages of LMAR is that it can be deployed parallel with small scale LLM on common personal devices given its modest hardware requirement. This configuration is particularly well-suited for applications such as enterprise search, medical knowledge retrieval, and offline technical support systems, where data privacy, low latency, and cost efficiency are critical.

\section{Discussion \& Future Work}
\label{Conclusion}
This work presents LMAR, a practical, efficient, and privacy-preserving framework for domain-adaptive retrieval in RAG systems. By leveraging LLM-guided contrastive learning and semantic chunking, LMAR addresses two key limitations in existing RAG pipelines: the domain knowledge gap of pre-trained embedding models and the ineffectiveness of standard chunking strategies in specialized contexts. In addition to improving retrieval performance and computational efficiency, LMAR offers a higher level of data privacy, making it suitable for RAG applications involving sensitive information such as patient records or proprietary business data. Unlike many commercial RAG systems that require uploading entire documents to centralized servers, LMAR minimizes data exposure through a piecewise interaction mechanism. During the synthetic data generation phase, only small, semantically coherent text segments such as individual paragraphs or short clusters are shared through the LLM API, thereby reducing the risk to disclose complete documents with confidential information. This presents a feasible solution for organizations with constrained computational resources and strict data privacy requirements to adopt the LMAR framework using locally hosted open-source LLMs.

Future efforts will focus on broadening its applicability while addressing the following theoretical and practical challenges. First, more advanced clustering algorithms can be explored to improve semantic grouping while reducing computational overhead. Second, LMAR currently focuses on Question-Answer tasks; future work will investigate task-oriented augmentation methods to support broader use cases such as contract analysis, clinical note summarization, or legal reasoning as well as adaptation to multi-modal QA with images.

\newpage
\bibliographystyle{unsrt}  
\bibliography{references} 

\newpage
\appendix
\section{LLM Reasoning Prompts}
\label{Sec:prompts}
\begin{center}
    \begin{tcolorbox}[
        colback=blue!5,
        colframe=blue!40!black,
        boxrule=2pt,
        arc=4pt,
        left=12pt,
        right=12pt,
        top=12pt,
        bottom=12pt,
        title={\Large\textbf{Triplet Labeling Prompt}},
        coltitle=white,
        colbacktitle=blue!80!black,
        fonttitle=\bfseries
    ]
    You will be provided with a triplet of texts. There is an anchor text and two candidate text paragraphs. Your task is to determine which one is more similar to the anchor text semantically.
            Each Candidate is labeled with special token |<1>| or |<2>|. Please first given reason for your decision and return corresponding special label at the end. \\

            Note:
            
            1. Please first given reason for your decision and return a corresponding special label in a dictionary format.
            
            2. First analyze on overall topic consistency, and then consider the context, entities, or event. \\

            3. If the two candidates are equally similar to the anchor text, return "Error" only.

            JSON format Example Out:
            {
            "Reason": "The reason for choosing the first candidate text |<1>| is that it described the potential solutions to the same problem as the anchor text",
            "Token": "|<1>|"
            }
    \end{tcolorbox}
\end{center}

\begin{center}
    \begin{tcolorbox}[
        colback=blue!5,
        colframe=blue!40!black,
        boxrule=2pt,
        arc=4pt,
        left=12pt,
        right=12pt,
        top=12pt,
        bottom=12pt,
        title={\Large\textbf{Cluster Description Generation Prompt}},
        coltitle=white,
        colbacktitle=blue!80!black,
        fonttitle=\bfseries
    ]
    You will be provided with a list of text paragraphs that belong to the same pre-defined cluster.\\
    Your task is to:\\ \\
    1.Write a concise summary that captures the **specific topic** shared by these paragraphs. The summary must include key entities such as **people, events, locations, or dates** mentioned in the text. Avoid vague or overly abstract descriptions.\\ \\
    2.Return the response strictly in this JSON format (including all punctuation and braces):\\ 
    \{ "description": "Clear and specific summary of the cluster"\} \\ \\
    Instructions: \\
    - Start your response with a left curly brace `\{` and end exactly at the closing brace `\}`.\\
    - Include double quotes** " ** around all keys and values.\\
    - Do **not** include any explanation or extra text.\\
    - Ensure the summary reflects the **actual shared content** of the cluster, not just generalizations.
    \end{tcolorbox}
\end{center}

\begin{center}
    \begin{tcolorbox}[
        colback=blue!5,
        colframe=blue!40!black,
        boxrule=2pt,
        arc=4pt,
        left=12pt,
        right=12pt,
        top=12pt,
        bottom=12pt,
        title={\Large\textbf{QA Pair Grading Prompt}},
        coltitle=white,
        colbacktitle=blue!80!black,
        fonttitle=\bfseries
    ]
    You are an AI assistant that evaluates the semantic alignment between a question and a candidate evidence sentence.\\

        You must return your evaluation **strictly** in this JSON format:\\

        \{"grade": float  // A score between 0.0 (completely irrelevant) and 1.0 (perfectly answers the question) \} \\ 

        \#\#\# Scoring Criteria: \\
        - 1.0: The sentence directly and clearly answers the question with specific, factual support.\\
        - 0.7–0.9: The sentence provides strong partial evidence or indirectly implies the answer.\\
        - 0.4–0.6: The sentence is related in topic but does not actually answer the question.\\
        - 0.1–0.3: The sentence is loosely related or only shares surface-level words.\\
        - 0.0: The sentence is completely irrelevant to the question or misleading.\\

        Return ONLY the JSON object. Do NOT explain your reasoning or add any extra text.
    \end{tcolorbox}
\end{center}

\begin{center}
    \begin{tcolorbox}[
        colback=blue!5,
        colframe=blue!40!black,
        boxrule=2pt,
        arc=4pt,
        left=12pt,
        right=12pt,
        top=12pt,
        bottom=12pt,
        title={\Large\textbf{QA Pair Generation Prompt}},
        coltitle=white,
        colbacktitle=blue!80!black,
        fonttitle=\bfseries
    ]
    You are an AI that generates structured JSON output.

    You will be given:\\
    - A brief summary describing the overall topic of a cluster of related text paragraphs.\\
    - A list of original paragraphs, each prepended by its **original ID from a larger corpus**.\\

    Your task is to generate a set of **fact-based questions**, and for each question, return a list of *evidence\_ids* corresponding to the corpus IDs of paragraphs that together provide enough information to answer it.\\

    Requirements:\\
    - Some questions should require only **one paragraph** to answer.\\
    - Some questions should require **two or more paragraphs** to answer (multi-hop).\\
    - Only generate questions that are clearly and completely answerable using the given evidence IDs.\\
    - You must use the exact corpus IDs provided for each paragraph (e.g., 51, 243, 921).\\
    ---\\
    Input Format:\\

    Summary:\\
    "A description of the cluster topic"\\

    Paragraphs:\\
    51: "Elvis performed a concert in Honolulu, Hawaii, in 1973."\\
    243: "The concert was broadcast live via satellite to over 40 countries."\\
    378: "This was the first time a concert was aired globally in real time."\\
    912: "The show raised money for cancer research."\\
    ---\\
    Output Format:
    \begin{verbatim}
    {
      "qa_pairs": [
        {
          "question": "Where did Elvis perform the 1973 concert, and how many
           countries received the broadcast?",
          "evidence_ids": [51, 243]
        },
        {
          "question": "Why was the 1973 concert considered historically significant?",
          "evidence_ids": [243, 378]
        },
        {
          "question": "What charitable cause benefited from the concert?",
          "evidence_ids": [912]
        }
      ]
    }
    \end{verbatim}
    --- \\
    Generation Rules: \\
    - Generate up to \{max\_question\_num\} QA pairs.\\
    - Each question must be **distinct and fully grounded** in the provided paragraphs.\\
    - Do **not** invent facts or ask speculative questions.\\
    - At least some questions should require combining two or more `evidence\_ids`.\\
    - Only use the **exact paragraph IDs** shown in the input — they correspond to entries in the full corpus.\\
    - Format your response as a valid JSON object, starting with `\{\{` and ending with `\}\}`.\\
    - Do NOT include any explanation or commentary outside the JSON object.\\
    - Double-check that each question is non-redundant and has valid evidence support.\\
    """format(max\_question\_num=max\_question\_num)
    
    \end{tcolorbox}
\end{center}

\section{Training Settings}
\label{Sec:Training Settings}
The training of all models follows a unified set of hyperparameters to ensure consistent comparisons across datasets and model variants. The values for learning rates, batch size, weight decay, and optimizer remain constant regardless of the base encoder model or whether clustering is used. These hyperparameters are detailed in Table~\ref{table:common-config-table}. 

\renewcommand{\thetable}{S\arabic{table}} 
\setcounter{table}{0} 

\begin{table}[H]
\centering
\caption{Common training configurations across all models and datasets.}
\label{table:common-config-table}
\begin{tabular}{ccccc}
\hline\hline
QE Init LR & Triplet Init LR & Batch Size & Optimizer \\
\hline
1e-06 & 1e-05 & 32 & AdamW \\
\hline\hline
\end{tabular}
\end{table}

The number of training epochs for each model variant is determined through early stopping based on the validation performance, using a held-out set excluded from the synthetic training data. Separate early stopping criteria are applied to the question-evidence pairs generation task and the triplet contrastive training task. The corresponding epoch numbers are summarized in Table~\ref{table:epochs-table} for each dataset and model configuration.


\begin{table}[H]
\centering
\caption{QE and Triplet training early stop epochs}
\label{table:epochs-table}
\begin{tabular}{lcc}
\hline\hline
Base Model & QE Ending Epochs & Triplet Ending Epochs \\
\hline\hline
\multicolumn{3}{l}{\textbf{Dataset: WikiQA}} \\
\hline
BERT No Cluster   & 14  & - \\
BERT              & 4 & 11 \\
\hline
\multicolumn{3}{l}{\textbf{Dataset: TechQA}} \\
\hline
BERT No Cluster   & 9  & - \\
BERT              & 19 & 17 \\
\hline
\multicolumn{3}{l}{\textbf{Dataset: PubmedQA}} \\
\hline
BERT No Cluster   & 9  & - \\
BERT              & 11 & 11 \\
\hline\hline
\end{tabular}
\end{table}

For WikiQA, the BERT model without clustering converges after 14 epochs on QE generation, while the clustered version requires only 4 epochs for QE and 11 epochs for Triplet training. In TechQA, the clustered BERT requires significantly more epochs (19) to converge for QE, compared to 9 for the no-cluster variant. On PubmedQA, QE training stops at 9 epochs (no cluster) and 11 epochs (with cluster). 

\end{document}